\documentclass[
  ,final            
  ,numberedheadings 
  ]
  {aipproc}

\layoutstyle{6x9}

\begin{document}

\title{Searches for New Physics with the ALICE Experiment in pp Collisions}


\author{A.~Dobrin, for the ALICE Collaboration}{
  address={Department of Physics, Division of Experimental High Energy Physics, Lund University, Box 188, SE-221 00, Lund, Sweden}
}

\begin{abstract}
The performance of the ALICE detector in searches for new heavy stable charged particles in pp collisions is discussed in this paper. Gluino R-hadron was chosen as an example of a candidate, and cross sections and kinematic properties were obtained from PYTHIA simulations for various gluino masses ($100-500~\mathrm{GeV/c}^2$). Detector response simulations for R-hadrons in the TPC acceptance ($|\eta|<1$) were performed for $100~\mathrm{GeV/c}^2$ R-hadron mass.
\end{abstract}

\maketitle

\section{Introduction}

The ALICE experiment~\cite{alice} at the CERN Large Hadron Collider (LHC)~\cite{lhc} is expected to record $10^9$ pp minimum bias events each year, so a relatively large cross-section to explore the physics landscape beyond the Standard Model (SM) is required. The new heavy stable charged particles ($m>100~\mathrm{GeV/c}^2$) predicted by some theoretical models~\cite{gmsb, guidice} provide good candidates for ALICE searches. Using a high precision tracking system and a time-of-flight detector for particle identification (PID) of slow particle, ALICE has two techniques for identification of heavy stable charged particles: the energy loss (dE/dx) method and the time-of-flight method.

In our study we focused on the Split Supersymmetry models~\cite{guidice}. These models predict a stable gluino or squark, but we only chose gluinos to form long lived particles (i.e. do not decay during their passage through the detector and undergo electromagnetic (EM) and/or strong interactions with matter). The stable gluinos hadronize into heavy (electrically charged or neutral) bound states called R-hadrons. The phenomenology of a stable gluino has been previously studied in Refs.~\cite{Kraan, mackeprang}. Different R-hadron types can be formed from a stable gluino: R-mesons ($\mathrm{\tilde{g} q \overline{q}}$), R-baryons ($\mathrm{\tilde{g} q q q}$), R-gluinoballs ($\mathrm{\tilde{g} g}$). R-mesons are supposedly the lightest hadron states and they are mass degenerated; R-baryons are also mass degenerated~\cite{Kraan}. The details of R-hadron interactions with normal matter are not very well predicted, but some features are common among the models. The gluino is considered as a heavy, non-interacting spectator, acting as a reservoir of kinetic energy and surrounded by a cloud of interacting quarks which determine the charge of the R-hadron. R-hadrons can have hadronic interactions with the detector and EM if charged. By exchanging quarks, R-hadrons may flip charge and baryon number in hadronic interactions. This is one of the signals that the other LHC experiments are relying on in their search for R-hadrons. Due to the very low material budget we only considered the non-hadronic interactions.

\section{The ALICE experiment}

\begin{figure}
  \includegraphics[keepaspectratio, width=0.7\columnwidth]{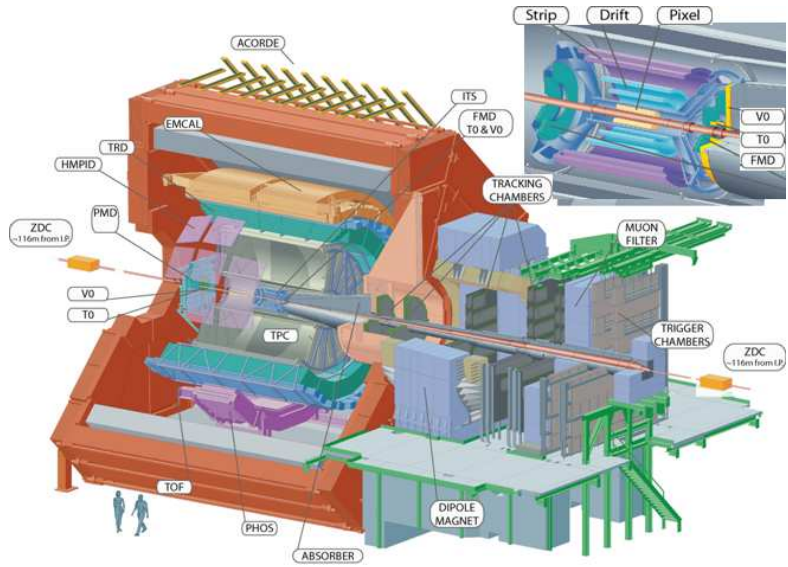}
  \caption{ALICE shematic layout.}
  \label{3d:alice}
\end{figure}

ALICE, the dedicated heavy ion experiment at the LHC, consists of a central barrel which measures hadrons, electrons, and photons, and a forward spectrometer for muons identification. The central part covers $|\eta|<0.9$ and is located in the L3-solenoid which provides a maximum magnetic field of 0.5 T. The barrel contains four detectors that cover the full azimuth: an Inner Tracking System (ITS) made of six planes of high-resolution silicon pixel (SPD), drift (SDD) and strip (SSD) detectors, a cylindrical Time Projection Chamber (TPC), a Transition Radiation Detector (TRD), and a Time-Of-Flight (TOF) detector. It also contains three detectors with limited acceptance: High-Momentum Particle Identification Detector (HMPID), a Photon Spectrometer (PHOS), and an Electromagnetic Calorimeter (EMCal). The Forward Muon Spectrometer covers $-4.0<\eta<-2.5$ and consists of absorbers, a large dipole magnet, and fourteen planes of tracking and triggering chambers. Several smaller detectors (ZDC, PMD, FMD, T0, V0) are used for triggering and global event characterization. An array of scintillators (ACORDE) on top of the L3 magnet is used to trigger on cosmic rays. A sketch of the apparatus is shown in Fig.~\ref{3d:alice}.

\section{Results and discussions}

Using PYTHIA 6.4.10~\cite{Torbjorn} we investigated gluino masses of 100, 200, 300, and 500~$\mathrm{GeV/c}^2$. After the gluinos are produced they are hadronized into color singlet bound states (R-hadrons). No $\mathrm{\tilde{g} \tilde{g}}$ bound states were considered here and thus the R-hadrons are always produced in pairs. The probability for a R-hadron to be charged is about 50$\%$. In order to see how many R-hadrons would have nearly full length tracks in the TPC, we used a cut in pseudorapidity $|\eta|<1$.

\begin{figure}
  \includegraphics[keepaspectratio, width=0.47\columnwidth]{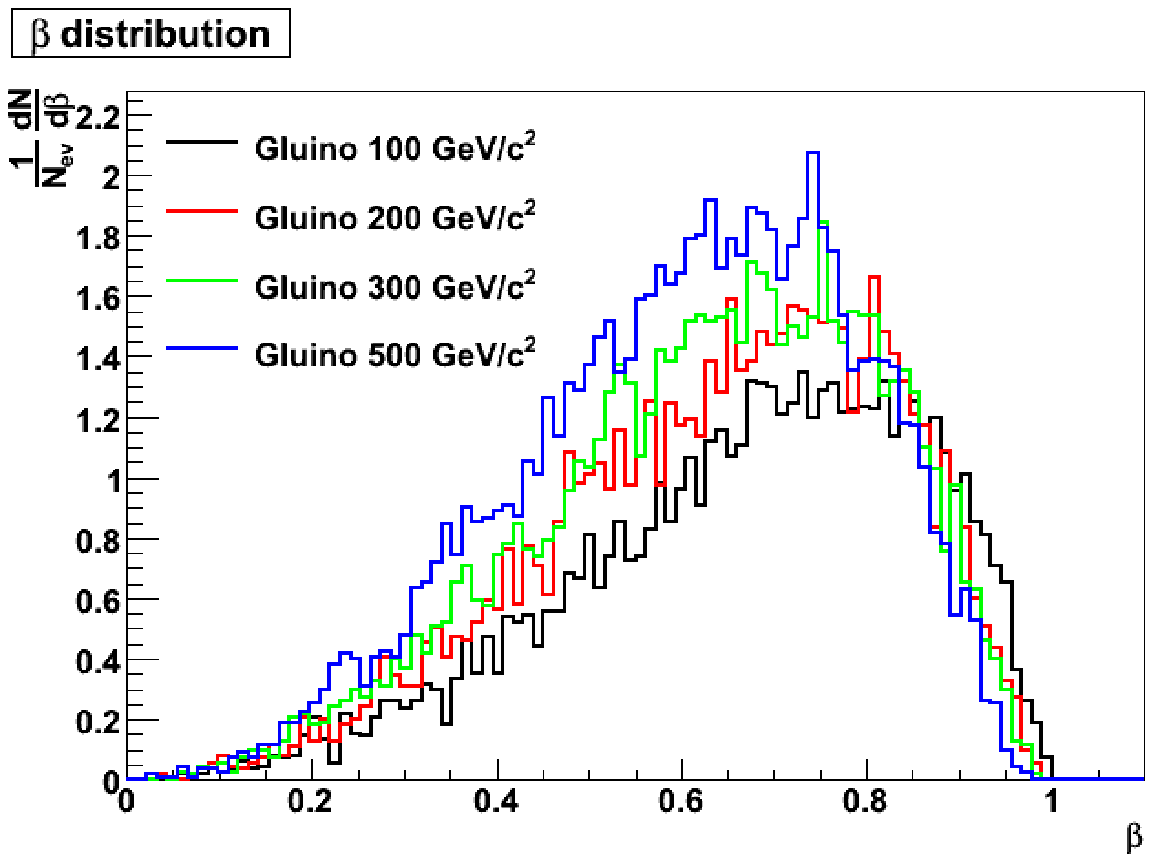}
  \includegraphics[keepaspectratio, width=0.47\columnwidth]{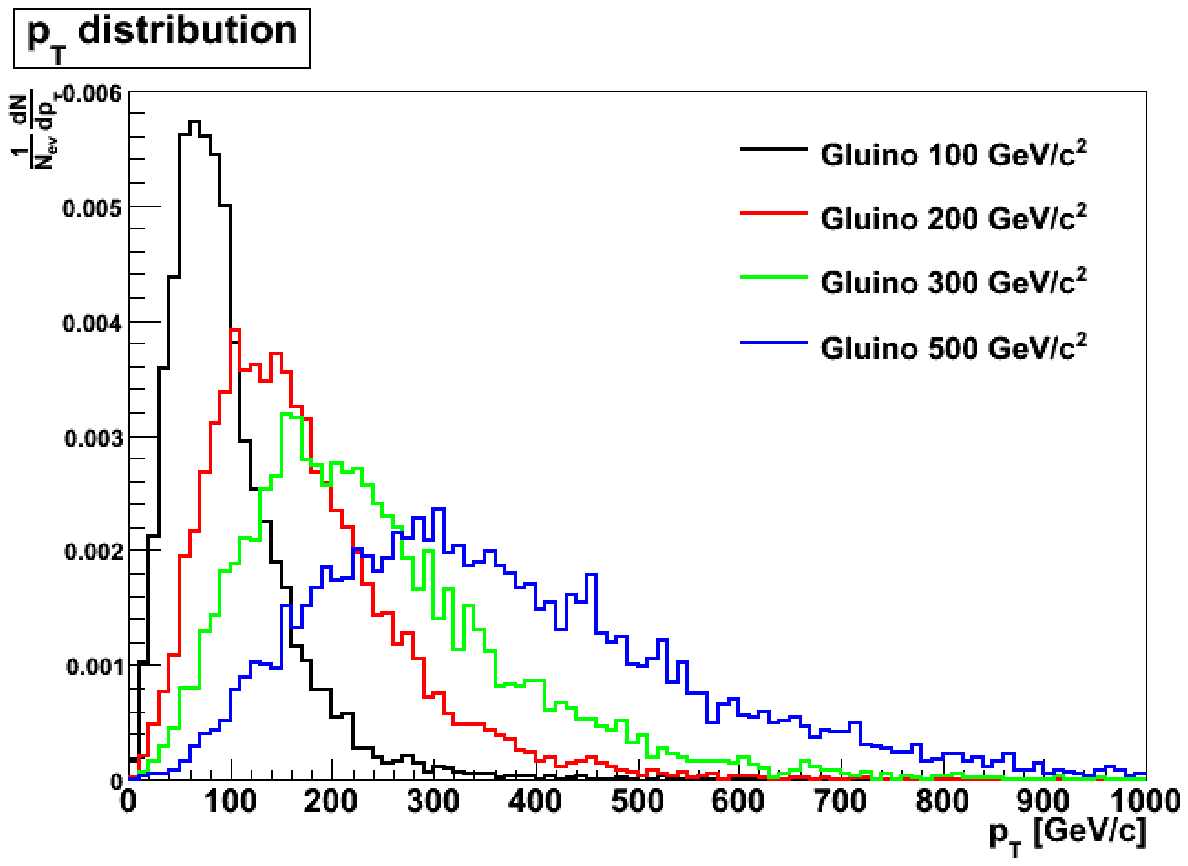}
  \caption{Kinematic properties of the R-hadrons for different values of $m_{\tilde{g}}$ in the TPC acceptance ($|\eta|<1$): $\beta$ distribution (left) and $p_{T}$ distribution (right).}
  \label{pythia:kin}
\end{figure}

The kinematics of the physics processes generated after the $\eta$ cut is shown in Fig.~\ref{pythia:kin}. The left plot shows the $\beta$ distribution ($\beta=\frac{p}{E}$), while the $p_T$ distribution is presented in the right one. Most of the R-hadrons have a $\beta$ between 0.3 and 0.9 with a mean around 0.7 (the centrally produced R-hadrons been slower than the forward produced ones). So R-hadrons are very different from SM particles with a similar momentum distribution: the slow R-hadrons will ionize like ``soft'' protons and will have a large time-of-flight. This difference can be exploited in order to identify heavy stable particles.

The detector response simulation results based on the PYTHIA output files will be presented next. We simulated and reconstructed charged R-hadrons with the mass fixed to $100~\mathrm{GeV/c}^2$, based on the PYTHIA simulation of $100~\mathrm{GeV/c}^2$ gluinos. The results of the R-hadron simulation were compared with the results of pion, proton, and muon simulations with exactly the same momentum distribution. The magnetic field used was 0.5 T. The reconstruction efficiency for ITS and TPC was found to be 83$\%$. When a TOF signal was required the reconstruction efficiency dropped to 65$\%$.

\begin{figure}
    \includegraphics[keepaspectratio, width=0.32\columnwidth]{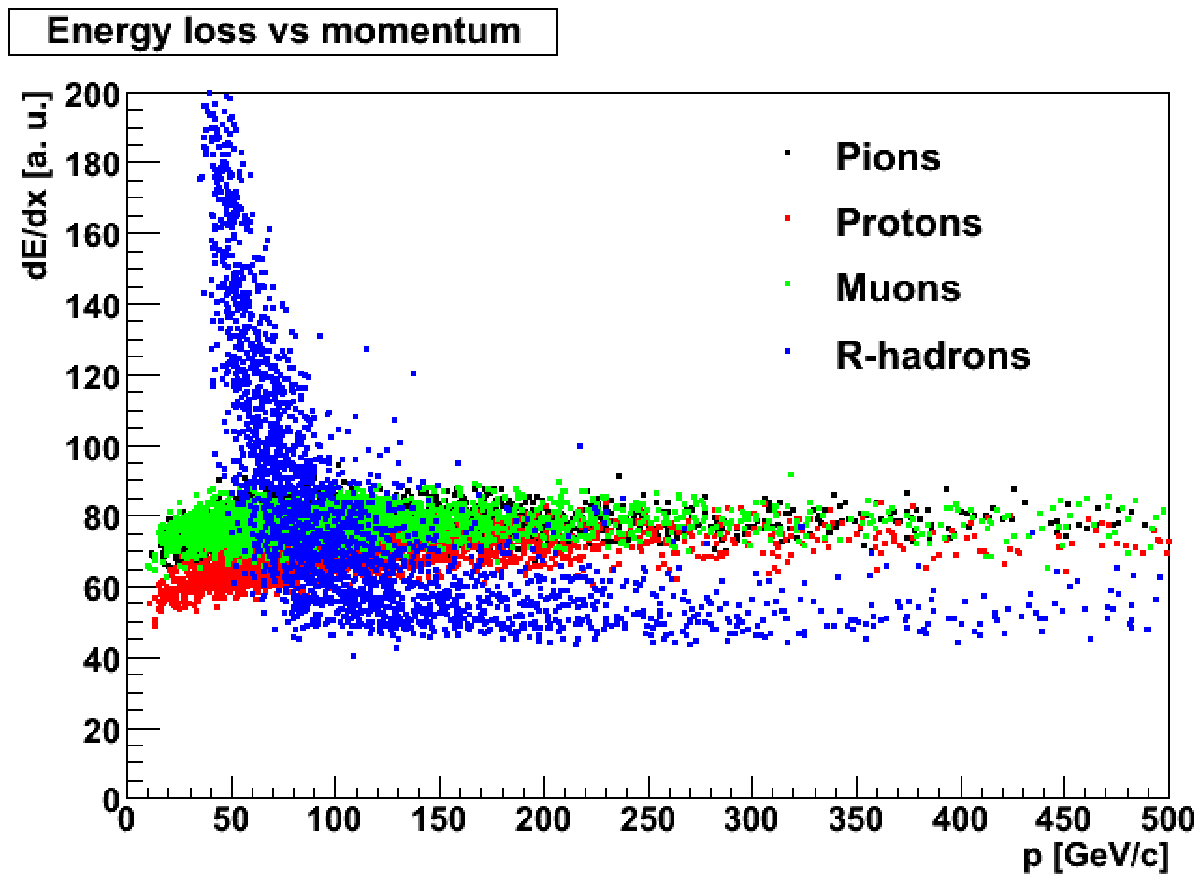}
    \includegraphics[keepaspectratio, width=0.32\columnwidth]{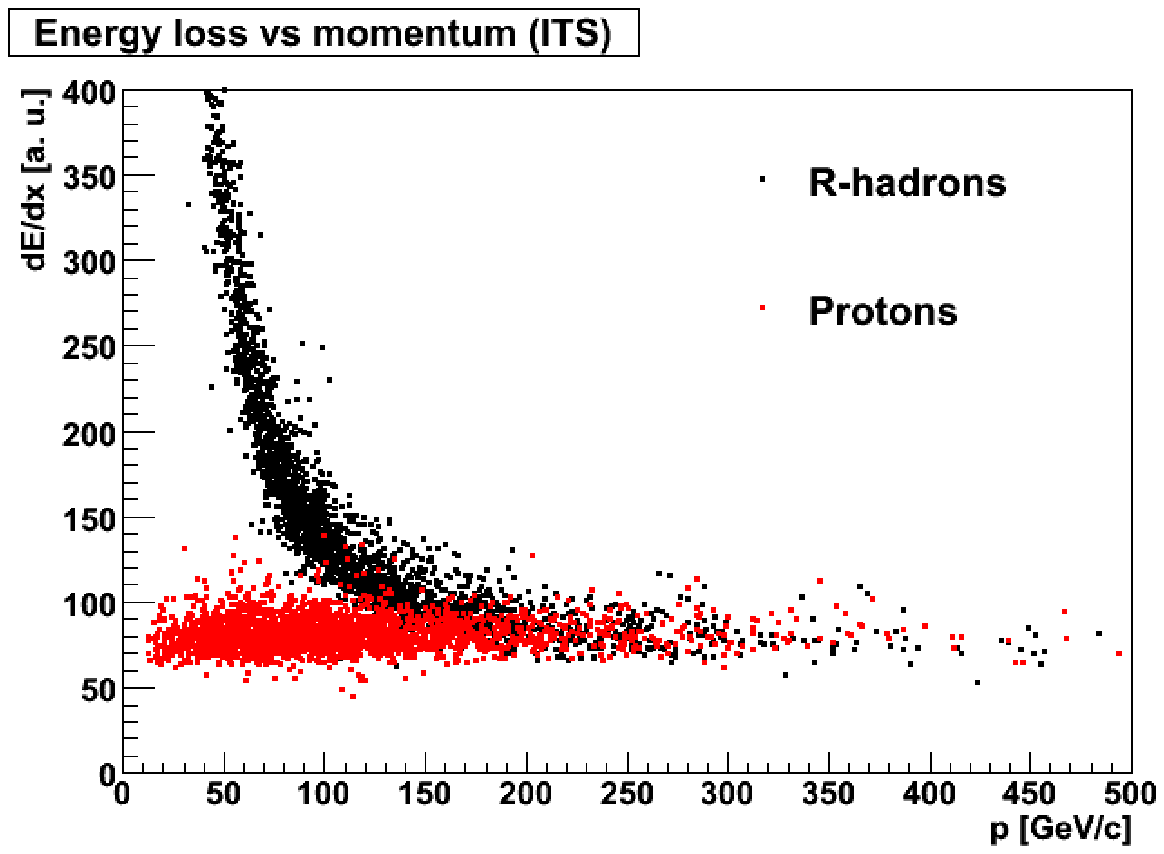}
    \includegraphics[keepaspectratio, width=0.32\columnwidth]{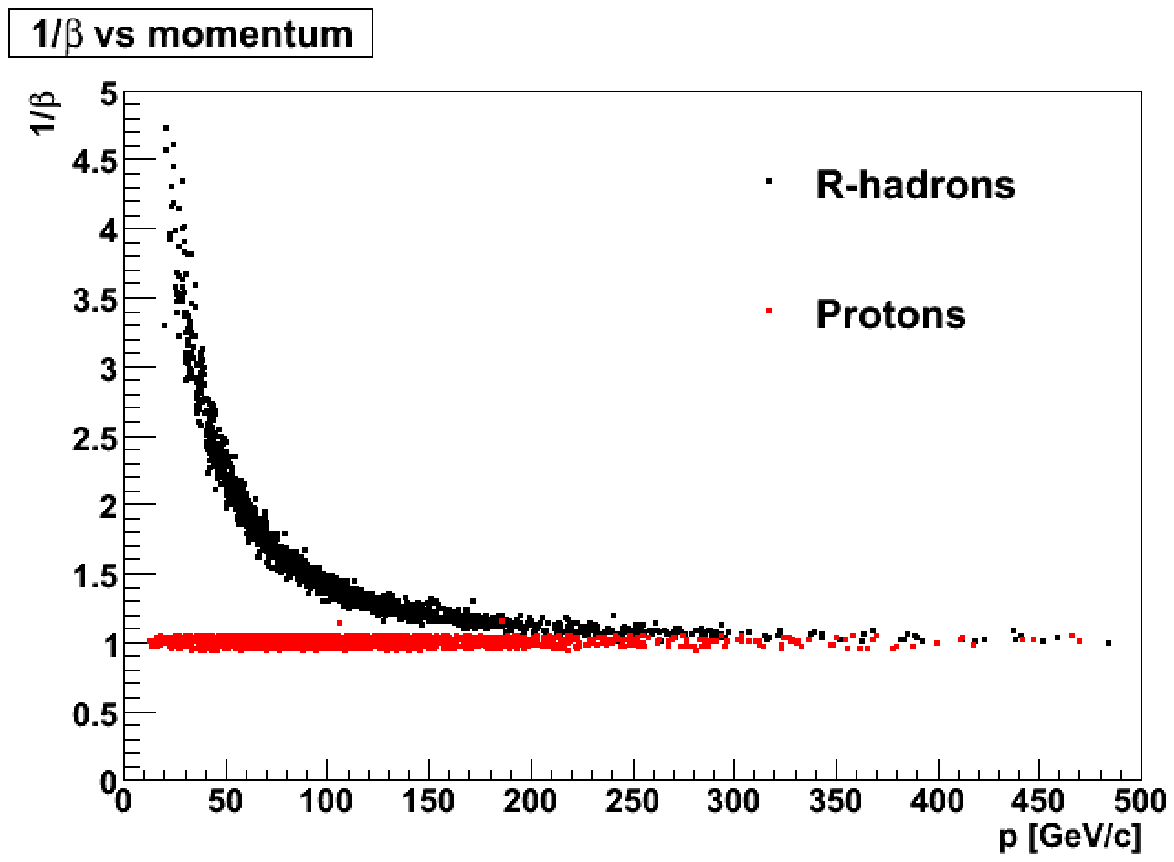}
  \caption{Detector responses for R-hadrons, pions, protons, and muons: dE/dx as a function of the reconstructed momentum for TPC (left) and ITS (middle) and the $1/\beta$ as a function of the reconstructed momentum for TOF (right).}
  \label{det:responses}
\end{figure}

The detector responses for R-hadrons, pions, protons, and muons are presented in Fig.~\ref{det:responses}. As we expected the dE/dx for R-hadrons lies in the $1/\beta^2$ region while the dE/dx for pions, protons, and muons are on the relativistic rise/plateau (for both ITS and TPC results). The TPC PID response is depicted in the left plot (the dE/dx distribution as a function of the reconstructed momentum for charge R-hadrons, pions, protons, and muons). The slow R-hadrons ($31\%$) are well separated from pions, protons, and muons ($p_{rec}>10~\mathrm{GeV/c}$, $dE/dx>100$). The remaining R-hadrons cannot be separated from other charged particles by the TPC on an event by event basis, but only statistically which is not likely to be feasible because of the low yield. The middle figure shows the dE/dx in ITS as a function of the reconstructed momentum for R-hadrons and protons. As can be seen from the figure, in the Silicon detector one can separate R-hadrons up to higher momenta ($p<100~\mathrm{GeV/c}$) because there is no relativistic rise. However, the dE/dx resolution is worse and might give a large background. From the TOF signal one can determine $\beta$. The right figure shows $1/\beta$ as a function of the reconstructed momentum for R-hadrons, and protons. As we expected the protons have a $1/\beta$ distribution around 1 (they are ultra-relativistic) while for R-hadrons $1/\beta$ first approaches 1 at very high momenta ($p>250~\mathrm{GeV/c}$). For $p_{rec}>10~\mathrm{GeV/c}$ and $1/\beta>1.04$, $99\%$ of R-hadrons have a 5$\sigma$ separation from protons. 

More details on the analysis and additional results can be found in Ref.~\cite{note}. Also, in order to enhance the R-hadron sample, two trigger scenarios are under study.

\section{Conclusions}
With the start-up of the LHC the discovery of physics beyond the Standard Model through heavy stable charged particles is a possibility also for ALICE due to its excellent particle identification capabilities. In order to have evidence for heavy stable charged hadrons we will probably need at least one ``golden event'' where we measure two back to back high $p_{T}$ tracks with dE/dx consistent with high masses in ITS, TPC, and a time-of-flight indicating that the particles are slow.

\bibliographystyle{aipproc}

\begin{thebibliography}{9}

\bibitem{alice}
ALICE Collaboration, K.~Aamodt et al., JINST \textbf{3}, S08002 (2008).

\bibitem{lhc}
L.~Evans and P.~Bryant (editors), JINST \textbf{3}, S08001 (2008).

\bibitem{gmsb}
G.~F.~Giudice and R.~Rattazzi, arXiv:hep-ph/9801271v2.

\bibitem{guidice}
N.~Arkani-Hamed, S.~Dimopoulos, G.~F.~Giudice and A.~Romanino, Nucl. Phys. \textbf{B 709} (2005) 3.

\bibitem{Kraan}
A.~C.~Kraan, \emph{Interactions of heavy stable hadronizing particles}, Eur. Phys. J. \textbf{C 37} (2004) 91.

\bibitem{mackeprang}
R. Mackeprang and A. Rizzi, Eur. Phys. J. \textbf{C 50} (2007) p. 353.

\bibitem{Torbjorn}
T. Sj\"ostrand, S. Mrenna and P. Skands \emph{PYTHIA 6.4 physics and manual}, JHEP \textbf{05} (2006) 026.

\bibitem{note}
 A.~Dobrin, P.~Christiansen, H-{\AA}.~Gustafsson, P.~Gros, A.~Oskarsson and E.~Stenlund, \emph{Search for Heavy Stable Charged Hadrons in pp Collisions with the ALICE Experiment}, ALICE-INT-2008-017 (public).

\end{thebibliography}

\end{document}